\documentclass[a4paper,english,english,aps]{revtex4}
\usepackage[T1]{fontenc}
\usepackage[latin1]{inputenc}

\makeatletter

\providecommand{\LyX}{L\kern-.1667em\lower.25em\hbox{Y}\kern-.125emX\@}

\usepackage{babel}
\makeatother
\begin{document}

\title{Linear and non-linear perturbations in dark energy models}

\author{Luca Amendola }

\address{INAF/Osservatorio Astronomico di Roma, \\
 Viale Frascati 33, 00040 Monte Porzio Catone (Roma), Italy\\
 \textit{amendola@mporzio.astro.it}}

\date{\today {}}

\begin{abstract}
I review the linear and second-order perturbation theory in dark energy
models with explicit interaction to matter in view of application
to $N$-body simulations and non-linear phenomena. Several new or
generalized results are obtained: the general equations for the linear
perturbation growth; an analytical expression for the bias induced
by a species-dependent interaction; the Yukawa correction to the gravitational
potential due to dark energy interaction; the second-order perturbation
equations in coupled dark energy and their Newtonian limit. I also
show that a density-dependent effective dark energy mass arises if
the dark energy coupling is varying.
\end{abstract}
\maketitle

\section{Introduction}

Dark energy is defined as a fluid distributed almost homogeneously
and whose potential energy dominates over kinetic energy. As such,
it can be observed mainly through large scale effects as those relating
to the cosmic expansion history and its linear fluctuations. Indeed,
the most weighty evidences in favor of dark energy come from the acceleration
of the universe as seen in the Hubble diagram of the supernovae Ia
\cite{rie} and on the angular size of the acoustic horizon on the
cosmic microwave background \cite{wmap}. Indications concerning the
growth of linear fluctuations, e.g. via the ISW effect \cite{coras,fosa}
are still very tentative, although the prospects from e.g. weak lensing
\cite{benabed} and Lyman-$\alpha $ clustering \cite{mandel,carva}
appear promising.

However, all these observables depend ultimately on dark energy only
through the expansion history $H(z)$ and the matter linear growth
function $D(z)$, where $z$ is the cosmological redshift. For instance,
luminosity distance in flat space is defined as\begin{equation}
d_{L}(z)=(1+z)\int _{0}^{z}\frac{dz}{H(z)},\label{eq:}\end{equation}
where \begin{equation}
H(z)=H_{0}\left[\Omega _{m}a^{-3}+\Omega _{\phi }a^{-3[1+W(a)]}+(1-\Omega _{m}-\Omega _{\phi })a^{-2}\right]^{1/2},\label{eq:}\end{equation}
where $a=(1+z)^{-1}$ is the scale factor, $\Omega _{i}$ denotes
the density at the present time of the $i$-th species and \begin{equation}
W(\alpha =\log a)=\frac{1}{\alpha }\int _{0}^{a}w_{\phi }(\alpha ')d\alpha ',\label{eq:}\end{equation}
$w_{\phi }(z)$ being the equation of state of dark energy. It is
clear that at any given redshift there will be different $w_{\phi }(z)$
that give indistinguishable $d_{L}(z)$ and that the degree of degeneration
will increase with redshift. Similar integrals of $H(z)$ enter the
definitions of angular-diameter distance and age, that will therefore
be subject to the same ambiguity. 

The linear growth function is a second observable quantity, in general
independent of $H(z)$. In a matter dominated epoch with present density
parameter $\Omega _{m0}$ and on sub-horizon scales it is given by
the solution of the perturbation equation\begin{equation}
D''+\frac{1}{2}\left(1+\frac{\mathcal{H}'}{\mathcal{H}}\right)D'-\frac{3\Omega _{m0}}{2}a^{-3}D=0\label{eq:}\end{equation}
(valid for uncoupled dark energy) where the prime denotes derivation
with respect to $\alpha =\log a$ and where we introduce the conformal
Hubble function $\mathcal{H}=d\alpha /d\tau $, $\tau $ being the
conformal time. The growth $D$ is therefore an independent probe
of dark energy: two models that give an identical $H(z)$ will in
general be distinguished by different $D(z)$.

In principle the degeneracy can be broken by a large number of observations
at different $z$ of $H(z)$ and/or $D(z)$. However, this is hardly
feasible, since real data is confined to {}``small'' ($z<5$) or
very large redshifts ($z\approx 1100$). Moreover, in most models
the dark energy component become subdominant at $z\gg 1$, so that
both $H(z)$ and $D(z)$ becomes rapidly insensitive to $w_{\phi }(z)$
(see \cite{kujat} for a detailed discussion on the practical observability
of $w_{\phi }(z)$ at large $z$).

It would be desirable therefore to add new observables, as the evolution
of the perturbations in the dark energy field itself, the second order
growth function or the full non-linear properties as obtained through
$N$-body simulations (see e.g. \cite{mainini,maccio,dolag}). In
this paper we review the general linear and non-linear perturbation
equations in dark energy models in order to provide the basic material
for the study of these additional observable quantities. Our dark
energy model is quite general: a scalar field with a generic potential
and an explicit varying species-dependent coupling to baryons and
dark matter. This cover most scalar field models presented in literature,
with the notable exception of models with non-standard kinetic terms
\cite{muk}.

In a subsequent paper we employ the formalism derived here to evaluate
the large scale skewness in coupled dark energy.

\section{Coupled dark energy}

Our dark energy model is characterized by a general potential $V(\phi )$
and general couplings $C_{i}(\phi )$ to matter. This class of models
include those motivated by string theory proposed in \cite{wet,ame3,gasp,extra1,chimento,agtu,carsten,others,others2,others3}.
The conservation equations with interacting terms for the field $\phi $
, cold dark matter ($c$), baryons ($b$ ) are:\begin{eqnarray}
T_{(c)\nu ;\mu }^{\mu } & = & -C_{c}(\phi )T_{(c)}\phi _{;\nu }\: ,\\
T_{(b)\nu ;\mu }^{\mu } & = & -C_{b}(\phi )T_{(b)}\phi _{;\nu }\: ,\label{eq:}\\
T_{(\phi )\nu ;\mu }^{\mu } & = & [C_{c}(\phi )T_{(c)}+C_{b}(\phi )T_{(b)}]\phi _{;\nu }\: ,\label{eq:cons}
\end{eqnarray}
 where the coupling functions $C_{b,c}(\phi )$ depend on the species,
as first proposed in \cite{dam}. Radiation (subscript $\gamma $)
remains uncoupled because it is traceless (models with coupling to
the electromagnetic field \cite{mb,em} or neutrinos \cite{neu} have
also been proposed). The standard Einstein equations are assumed to
hold. This coupling form is derived, through a conformal transformation,
from a Brans-Dicke gravity with species-dependent interaction \cite{dam,ame1}.
There are strong limits on the baryon coupling \cite{key-69} and
relatively looser ones on the dark matter coupling from either astrophysics
\cite{grad,maccio} or cosmology \cite{amen02}; here, however, we
disregard any constraint and leave the couplings as free functions.
In a FRW metric the equations (\ref{eq:cons}) plus the Friedmann
equation read: \begin{eqnarray}
\ddot{\phi }+3H\dot{\phi }+V_{,\phi } & = & \sqrt{2/3}\kappa (\beta _{c}\rho _{c}+\beta _{b}\rho _{b}),\nonumber \\
\dot{\rho }_{c}+3H\rho _{c} & = & -\sqrt{2/3}\kappa \beta _{c}\rho _{c}\dot{\phi },\nonumber \\
\dot{\rho }_{b}+3H\rho _{b} & = & -\sqrt{2/3}\kappa \beta _{b}\rho _{b}\dot{\phi },\label{sys}\\
\dot{\rho }_{\gamma }+4H\rho _{\gamma } & = & 0,\nonumber \\
3H^{2} & = & \kappa ^{2}\left(\rho _{\gamma }+\rho _{c}+\rho _{b}+\rho _{\phi }\right),\nonumber 
\end{eqnarray}
 where $\kappa ^{2}=8\pi G$, $\beta _{c}=C_{c}\sqrt{\frac{3}{2\kappa ^{2}}},$
$\beta _{b}=C_{b}\sqrt{\frac{3}{2\kappa ^{2}}},$ $H=\dot{a}/a$.
The matter conservation equations can be integrated out:\begin{equation}
\rho _{c,b}=\rho (0)_{c,b}a^{-3}\exp [-\sqrt{\frac{2}{3}}\kappa \int \beta _{c,b}(\phi )d\phi ].\label{eq:consmatt}\end{equation}
This shows one of the basic properties of dark energy interactions:
matter density appears to be non-conserved. For as concerns the potential
$V(\phi )$ here we write in all generality\begin{equation}
V(\phi )=Ae^{-\kappa \sqrt{2/3}\mu f(\phi )\phi }\: ,\label{eq:potential}\end{equation}
where $\mu $ is a dimensionless constant. The exponential case studied
in \cite{wet,ame3} corresponds therefore to $f=1$ , a constant potential
as in \cite{carsten} to $\mu =0$, and the power law $V\sim \phi ^{-n}$
to $f(\phi )=\sqrt{(3/2)}n\log \phi /(\kappa \mu \phi )$. We will
use also the definitions\begin{eqnarray}
\frac{dV}{d\phi } & = & -\sqrt{\frac{2}{3}}\kappa \mu f_{1}V\: ,\label{eq:}\\
\frac{d^{2}V}{d\phi ^{2}} & = & \frac{2}{3}\kappa ^{2}\mu ^{2}f_{2}V\: ,\label{eq:}
\end{eqnarray}
where\begin{eqnarray}
f_{1} & = & \frac{df}{d\phi }\phi +f\: ,\label{eq:}\\
f_{2} & = & f_{1}^{2}-\frac{\sqrt{3/2}}{\kappa \mu }\frac{df_{1}}{d\phi }\: .\label{eq:}
\end{eqnarray}
The simple exponential case reduces then to $f=f_{1}=f_{2}=1$. We'll
need later on higher derivatives of $V$ so we give the general rule:\[
V_{,\phi }^{(n)}=(-1)^{n}\left(\frac{2}{3}\right)^{n/2}\kappa ^{n}\mu ^{n}f_{n}V\: ,\]
where $f_{0}=1$ and\begin{equation}
f_{n}=f_{n-1}f_{1}-f'_{n-1}/(\sqrt{\frac{2}{3}}\kappa \mu )\: .\label{eq:fn}\end{equation}

The system (\ref{sys}) is best studied in the new variables \cite{cop,ame3}
\begin{equation}
x=\kappa \frac{\phi '}{\sqrt{6}},\quad y=\frac{\kappa }{H}\sqrt{\frac{U}{3}},\quad z=\frac{\kappa }{H}\sqrt{\frac{\rho _{\gamma }}{3},}\quad v=\frac{\kappa }{H}\sqrt{\frac{\rho _{b}}{3},}\end{equation}
 and the time variable $\alpha =\log a$. Then we obtain \begin{eqnarray}
x^{\prime } & = & \left(\frac{z^{\prime }}{z}-1\right)x-\mu f_{1}y^{2}+\beta _{c}(1-x^{2}-y^{2}-v^{2}-z^{2})+\beta _{b}v^{2},\nonumber \\
y^{\prime } & = & \mu f_{1}xy+y\left(2+\frac{z^{\prime }}{z}\right),\nonumber \\
z^{\prime } & = & -\frac{z}{2}\left(1-3x^{2}+3y^{2}-z^{2}\right),\label{eq:sys2}\\
v^{\prime } & = & -\frac{v}{2}\left(3\beta _{b}x-3x^{2}+3y^{2}-z^{2}\right).\nonumber 
\end{eqnarray}
 The CDM energy density parameter is obviously $\Omega _{c}=1-x^{2}-y^{2}-z^{2}-v^{2}$
while we also have $\Omega _{\phi }=x^{2}+y^{2}$, $\Omega _{\gamma }=z^{2}$
and $\Omega _{b}=v^{2}$. The system is subject to the condition $x^{2}+y^{2}+v^{2}+z^{2}\leq 1$.
To close the system one needs also the relation $f_{1}(y,H)$ and
the Friedman equation\begin{equation}
\frac{H'}{H}=-\frac{1}{2}\left(3+3x^{2}-3y^{2}+z^{2}\right).\label{eq:ht}\end{equation}
In the perturbation calculations we will use the conformal Hubble
function, $\mathcal{H}=aH$, and we will need the following relation\begin{equation}
\frac{\mathcal{H}'}{\mathcal{H}}=1+\frac{H'}{H}\label{eq:hubconf}\end{equation}

\section{Linear perturbations}

Aim of this section is to write down the linear perturbation equations
for a combination of fluid components with general equations of state
$p_{i}=w_{i}(\rho )\rho _{i}$ and a scalar field with a general potential
$V(\phi )$ and coupling $\beta _{i}(\phi )$ to the fluids. Some
of this material has been already published (e.g. in \cite{ame-bias,carsten})
but not at this level of generality. We choose the longitudinal gauge
\begin{equation}
ds^{2}=a^{2}[-(1+2\Psi )d\tau ^{2}+(1-2\Phi )dx_{i}dx^{i}],\label{eq:}\end{equation}
where $\tau $ is the conformal time. We will write the equations
in the time variable $\alpha =\log a$, so that our metric is effectively
\begin{equation}
ds^{2}=e^{2\alpha }[-(1+2\Psi )\frac{d\alpha ^{2}}{\mathcal{H}^{2}}+(1-2\Phi )dx_{i}dx^{i}].\label{eq:}\end{equation}
It is well-known that in absence of anisotropic stress $\Phi =\Psi $,
so we adopt this simplification from the start. We define the perturbation
variables \begin{equation}
\delta =\delta \rho /\rho ,\quad \varphi =\kappa \delta \phi /\sqrt{6},\quad v_{i}=\frac{adx_{i}}{\mathcal{H}dt},\quad \nabla _{i}v_{i}=\theta .\label{varia}\end{equation}
Repeated indexes mean summation and all spatial derivatives are with
respect to the comoving coordinates $x_{i}$. We define also the dark
energy mass \begin{equation}
m_{\phi }^{2}=\frac{d^{2}V}{d\phi ^{2}}\: ,\label{eq:}\end{equation}
with its dimensionless version\begin{equation}
\hat{m}_{\phi }^{2}=\frac{m_{\phi }^{2}}{H^{2}}=2\mu ^{2}y^{2}f_{2}\: .\label{eq:mass}\end{equation}
Perturbing the Einstein equations and the conservation equations we
obtain the linear perturbations below. We introduce the scale $\lambda $:
in real space we interpret $\lambda ^{-2}$ as the operator $-\mathcal{H}^{-2}\nabla ^{2}$;
in Fourier space, $\lambda =\mathcal{H}/k$. In this way the following
equations can be read equivalently in real or Fourier space. Note
also that $\beta '=d\beta /d\alpha =\phi 'd\beta /d\phi $.

The perturbation equations for a generic equation of state $p=w(\rho )\rho $,
which includes unified models like the Chaplygin gas \cite{chap}
or phase transition models \cite{bruce}, are

\textbf{Generic fluid component} (equation of state $p=w(\rho )\rho $)\begin{eqnarray}
\delta ' & = & -(w+1)\theta +3(1+w)\Phi '-2(1-3w)(\beta \varphi '+\beta '\varphi )-3w_{,\rho }\delta (1-2x\beta ),\\
\theta ' & = & -[(1-3w)(1-2x\beta )-w_{,\rho }A(w)+\frac{\mathcal{H}'}{\mathcal{H}}]\theta +\frac{w+w_{,\rho }}{w+1}\lambda ^{-2}\delta +2\beta \frac{3w-1}{w+1}\lambda ^{-2}\varphi +(1+w)\lambda ^{-2}\Psi ,\label{genw}
\end{eqnarray}
where $w_{,\rho }\equiv dw/d\log \rho ,$ and\begin{equation}
A(w)=3+2x\beta (1-3w)/(1+w).\label{eq:}\end{equation}
Note that the sound speed is $c_{s}^{2}\equiv dp/d\rho =w+w_{,\rho }$.
The equations for the scalar field coupled to several fluids with
equations of state $p_{i}=w_{i}(\rho _{i})\rho _{i}$ and the metric
equations, respectively, are

\textbf{Scalar field}\begin{eqnarray}
\varphi ''+\big (2+\frac{{\mathcal{H}}'}{\mathcal{H}}\big )\varphi '+(\lambda ^{-2}+\hat{m}_{\phi }^{2})\varphi -4\Phi 'x-2y^{2}\mu f_{1}\Phi  & = & \label{pert4w}\\
\sum _{i}\beta _{i}[1-3w_{i}-3w_{i,\rho }]\Omega _{i}\delta _{i}+2\sum _{i}\beta _{i}\Omega _{i}\Phi +\sum _{i}(1-3w_{i})\frac{\varphi }{x}\beta _{i}'\Omega _{i} & , & 
\end{eqnarray}

\textbf{Metric }

\begin{eqnarray}
 & \Phi  & =\frac{-3\lambda ^{2}[6x\varphi +2x\varphi '-2y^{2}\mu f_{1}\varphi +\sum \Omega _{i}(\delta _{i}+3(w_{i}+1)\lambda ^{2}\theta _{i})]}{2(1-3\lambda ^{2}(x^{2}+2y^{2}))},\label{pert5}\\
 & \Phi ' & =\frac{1}{2}[2(3x\varphi -\Phi )+\lambda ^{2}\sum 3(w_{i}+1)\theta _{i}\Omega _{i}].
\end{eqnarray}
From these general equations one can derive the following equations
for the three perfect fluid components, CDM ($w\approx 0\approx c_{s}$),
baryons ($w\approx 0,$ but $c_{s}^{2}$ non-negligible at small scales),
radiation ($w=1/3$).

\textbf{CDM }

\begin{eqnarray}
\delta '_{c} & = & -\theta _{c}+3\Phi '-2\beta _{c}\varphi '-2\beta _{c}'\varphi \: ,\\
\theta '_{c} & = & -\left(1+\frac{\mathcal{H}'}{\mathcal{H}}-2\beta _{c}x\right)\theta _{c}+\lambda ^{-2}(\Phi -2\beta _{c}\varphi )\: ,\label{cdm-n}
\end{eqnarray}

\textbf{Radiation \begin{eqnarray}
\delta '_{\gamma } & = & -\frac{4}{3}\theta _{\gamma }+4\Phi '\: ,\\
\theta '_{\gamma } & = & -\frac{{\mathcal{H}}'}{\mathcal{H}}\theta _{\gamma }+\frac{1}{4}\lambda ^{-2}\delta _{\gamma }+\lambda ^{-2}\Phi \: ,\label{rad-n}
\end{eqnarray}
}

\textbf{Baryons} \begin{eqnarray}
\delta '_{b} & = & -\theta _{b}+3\Phi '-2\beta _{b}\varphi '-2\beta _{b}'\varphi \: ,\\
\theta '_{b} & = & -\left(1+\frac{{\mathcal{H}}'}{\mathcal{H}}-2\beta _{b}x\right)\theta _{b}+c_{s}^{2}\lambda ^{-2}\delta +\lambda ^{-2}(\Phi -2\beta _{b}\varphi )\: \label{bar-n}
\end{eqnarray}

(to the equation for $\theta _{b}'$ one should add the standard term
describing momentum exchange with photons due to Thomson scattering,
see \cite{ma}).

\textbf{Scalar field\begin{eqnarray}
\varphi ''+\big (2+\frac{{\mathcal{H}}'}{\mathcal{H}}\big )\varphi '+(\lambda ^{-2}+\hat{m}_{\phi }^{2})\varphi -4\Phi 'x-2y^{2}\mu f_{1}\Phi  & = & \nonumber \\
\beta _{c}\Omega _{c}(\delta _{c}+2\Phi )+\beta _{b}\Omega _{b}(\delta _{b}+2\Phi )+\frac{\varphi }{x}(\Omega _{c}\beta _{c}'+\Omega _{b}\beta '_{b})\: , &  & \label{pert4}
\end{eqnarray}
}

Let us derive now the Newtonian limit (small scales, $\lambda \ll 1$).
The gravitational potential is \begin{eqnarray}
\Phi  & = & -\frac{3}{2}\lambda ^{2}(\sum \Omega _{i}\delta _{i}+6x\varphi +2x\varphi '-2y^{2}\mu f_{1}\varphi )\: ,\label{eq:poten}\\
\Phi ' & = & 3x\varphi -\Phi .\label{eq:poten2}
\end{eqnarray}
 Inserting (\ref{eq:poten2}) in (\ref{pert4}) we obtain \begin{equation}
\varphi ''+\big (2+\frac{{\mathcal{H}}'}{\mathcal{H}}\big )\varphi '+\varphi (\lambda ^{-2}+\hat{m}_{\phi }^{2}-12x-\frac{1}{x}\sum \Omega _{i}\beta '_{i})+\Phi (4x-2y^{2}\mu f_{1}-2\sum \Omega _{i}\beta _{i})=\sum \beta _{i}\Omega _{i}\delta _{i}\: ,\label{phi2}\end{equation}
 where the sum is on the coupled components (here baryons and dark
matter). It is interesting to observe that the terms in $\beta _{i}'$
contribute to the equation as effective masses. If for instance $\beta _{i}=\beta _{0}e^{\sqrt{2/3}\kappa \beta _{1}\phi }$
then we can define a {}``coupling mass''\begin{equation}
\hat{m}_{\beta _{i}}^{2}\equiv \frac{\Omega _{i}\beta '_{i}}{x}=2\Omega _{i}\beta _{i}\beta _{1}\: .\label{eq:}\end{equation}
The Newtonian limit in Eq. (\ref{phi2}) amounts to neglecting the
derivatives of $\varphi $ (because we can average out the rapid oscillations
of $\varphi $) and the metric potential $\Phi $ (which is proportional
to $\lambda ^{2}$). We can neglect also the term $12x\varphi $ since
$|x|\le 1$ is much smaller that $\lambda ^{-2}$. Finally, we neglect
here also $\hat{m}_{\phi }^{2}$ and $\hat{m}_{\beta _{i}}^{2}$ with
respect to $\lambda ^{-2}$ since otherwise the dark energy would
cluster on astrophysical scales and would reduce to a form of massive
dark matter \cite{guzman}. In section V., however, we remove this
approximation. From (\ref{phi2}) finally we obtain\textbf{\begin{equation}
\varphi \approx \lambda ^{2}(\beta _{c}\Omega _{c}\delta _{c}+\beta _{b}\Omega _{b}\delta _{b}).\label{phiappr-n}\end{equation}
}Since $\varphi $ is of order $\lambda ^{2}$, Eq. (\ref{eq:poten})
reduces to the usual Poisson equation (hereafter we neglect radiation)
\begin{equation}
\Phi =-\frac{3}{2}\lambda ^{2}(\Omega _{b}\delta _{b}+\Omega _{c}\delta _{c})\: .\label{eq:poisnewt}\end{equation}
Now, if we substitute in (\ref{cdm-n}) we can define a new potential
acting on dark matter\begin{equation}
\Phi _{c}=\Phi -2\beta _{c}\varphi =-\frac{3}{2}\lambda ^{2}\Omega _{b}\delta _{b}(1+\frac{4}{3}\beta _{b}\beta _{c})-\frac{3}{2}\lambda ^{2}\Omega _{c}\delta _{c}(1+\frac{4}{3}\beta _{c}^{2})\: .\label{eq:newpot}\end{equation}
In real space, this equation becomes\begin{equation}
\nabla ^{2}\Phi _{c}=4\pi G_{bc}\rho _{b}\delta _{b}+4\pi G_{cc}\rho _{c}\delta _{c}\: ,\label{eq:newpois}\end{equation}
where I restore the gravitational constant and define\begin{equation}
G_{ij}=G\gamma _{ij}\: ,\quad \gamma _{ij}\equiv 1+4\beta _{i}\beta _{j}/3\: ,\label{eq:geng}\end{equation}
so that $G_{bc}=G(1+4\beta _{b}\beta _{c}/3)$ and $G_{cc}=G(1+4\beta _{c}^{2}/3)$.
Analogous equations hold for the baryon force equation (\ref{bar-n}).
Therefore the Newtonian linear equations for dark matter and baryons
in coupled dark energy are\begin{eqnarray}
\delta '_{c} & = & -\theta _{c}\: ,\\
\theta '_{c} & = & -\left(1+\frac{\mathcal{H}'}{\mathcal{H}}-2\beta _{c}x\right)\theta _{c}-\mathcal{H}^{-2}\nabla ^{2}\Phi _{c}\: ,\label{cdm-n-2}\\
\delta '_{b} & = & -\theta _{b}\: ,\\
\theta '_{b} & = & -\left(1+\frac{\mathcal{H}'}{\mathcal{H}}-2\beta _{b}x\right)\theta _{b}-\mathcal{H}^{-2}\nabla ^{2}\Phi _{b}\: ,\\
\nabla ^{2}\Phi _{c} & = & 4\pi G_{bc}\rho _{b}\delta _{b}+4\pi G_{cc}\rho _{c}\delta _{c}\: ,\\
\nabla ^{2}\Phi _{b} & = & 4\pi G_{bb}\rho _{b}\delta _{b}+4\pi G_{bc}\rho _{c}\delta _{c}\: .
\end{eqnarray}
The $\beta '\varphi $ terms in the $\delta '$ equations have been
dropped because $\varphi $ is of order $\lambda ^{2}$. Deriving
the $\delta _{c}'$ equations we obtain \begin{equation}
\delta ''_{c}+\left(1+\frac{\mathcal{H}'}{\mathcal{H}}-2\beta _{c}x\right)\delta _{c}'-\frac{3}{2}(\gamma _{cc}\delta _{c}\Omega _{c}+\gamma _{bc}\delta _{b}\Omega _{b})=0,\label{deltacsimp}\end{equation}
 and similarly for $\delta '_{b}$\begin{equation}
\delta ''_{b}+\left(1+\frac{\mathcal{H}'}{\mathcal{H}}-2\beta _{b}x\right)\delta _{b}'-\frac{3}{2}(\gamma _{bc}\delta _{c}\Omega _{c}+\gamma _{bb}\delta _{b}\Omega _{b})=0.\label{deltabsimp}\end{equation}
These equations generalize previous results \cite{ame-bias} because
are valid also for non constant $\beta $ (provided $\hat{m}_{\beta }^{2}\ll \lambda ^{-2}$). 

It is clear that since baryons and dark matter obey different equations,
they will develop a bias already at the linear level. A simple result
can be obtained in the case in which one component dominates. Assuming
$\Omega _{b}\ll \Omega _{c}$, in fact, the baryon solution will be
forced by the dominating CDM to follow asymptotically its evolution.
Putting then $\delta _{c}\sim e^{\int m(\alpha )d\alpha }$ and $\delta _{b}=b\delta _{c}$
with $b=const.$ we obtain the coupled equations \begin{eqnarray}
m'+m^{2}+\left(1+\frac{{\mathcal{H}}'}{\mathcal{H}}-2\beta _{c}x\right)m-\frac{3}{2}\gamma _{cc}\Omega _{c} & = & 0\: ,\nonumber \\
m'+m^{2}+\left(1+\frac{\mathcal{H}'}{\mathcal{H}}-2\beta _{b}x\right)m-\frac{3}{2b}\gamma _{bc}\Omega _{c} & = & 0\: ,\label{eq:}
\end{eqnarray}
from which by subtraction \begin{equation}
b=\frac{3\gamma _{bc}\Omega _{c}}{3\gamma _{cc}\Omega _{c}+4(\beta _{c}-\beta _{b})xm}\: .\label{eq:bias-gen}\end{equation}
Notice that all terms on the right hand side are in general function
of time. This shows that a linear bias of gravitational nature develops
whenever $\beta _{c}\not =\beta _{b}$. This bias extends to all Newtonian
scales and therefore is distinguishable from the hydrodynamical or
non-linear bias that takes place in collapsed objects.

Further insight can be gained when these equations have constant coefficients,
i.e. when $w_{\phi }$ and $\Omega _{c,b}$ are constant (in \cite{ame3}
we denoted these cases as stationary solutions) . As shown in \cite{ame3}
this is realized on the critical points of a field governed by an
exponential potential and a constant coupling $\beta $. In this case
is convenient to define a total equation of state \begin{equation}
w_{e}=p_{tot}/\rho _{tot}=x^{2}-y^{2},\label{eq:}\end{equation}
instead of $w_{\phi }=p_{\phi }/\rho _{\phi }$. Neglecting the baryons
the relation is simply \[
w_{e}=(1-\Omega _{c})w_{\phi },\]
so that now \begin{equation}
\frac{\mathcal{H}'}{\mathcal{H}}=-\frac{1}{2}\left[1+3w_{e}\right].\label{eq:hphc}\end{equation}
The scale factor in this case grows as $a\sim \tau ^{p/(1-p)}\sim t^{p}$
where $p=2/[3(w_{e}+1)]$. The solutions are $\delta _{c}=a^{m_{\pm }}$
and $\delta _{b}=ba^{m_{\pm }}$ where $b$ is as in (\ref{eq:bias-gen})
and, again neglecting the baryons, i.e. for $\Omega _{b}\ll \Omega _{c}$
\cite{ame-bias}\begin{equation}
m_{\pm }=\frac{1}{4}\left(-1+3w_{e}+4\beta _{c}x\pm \Delta \right),\label{eq:}\end{equation}
where $\Delta ^{2}=24\gamma _{cc}\Omega _{c}+(-1+3w_{e}+4\beta _{c}x)^{2}$.
In this case then $m$ and $b$ are constant. The scalar field solution
is\begin{equation}
\varphi \approx \lambda ^{2}\beta _{c}\delta _{c}\Omega _{c}=H_{0}^{2}k^{-2}a^{2(p-1)/p}\beta _{c}\delta _{c}\Omega _{c}.\label{eq:phi-growth}\end{equation}
 For small wavelengths $\varphi $ (which here is proportional to
$\delta \rho _{\phi }/\rho _{\phi }$ ) is always much smaller than
$\delta _{c},\delta _{b}$ at the present time, unless of course $\beta _{c}$
is exceedingly large. It is interesting to observe that $\varphi $
could outgrow the matter perturbations in the future in an accelerated
epoch, i.e. if $p>1$ and if $\Omega _{c}$ does not vanish.

\section{Synchronous gauge }

Since most Boltzmann codes in CMB are implemented via the synchronous
gauge we give here the relevant equations in this gauge.

\textbf{Generic fluid component} (equation of state $p=w(\rho )\rho $
)\begin{eqnarray}
\delta ' & = & -(w+1)\theta -\frac{1}{2}(w+1)h'-2(1-3w)(\beta \varphi '+\beta '\varphi )-3w_{,\rho }\delta (1-2x\beta )\: ,\label{eq:dw}\\
\theta ' & = & -\left[(1-3w)(1-2\beta x)-w_{,\rho }A(w)+\frac{\mathcal{H}'}{\mathcal{H}}\right]\theta +\frac{w+w_{,\rho }}{w+1}\lambda ^{-2}\delta +2\beta \frac{3w-1}{w+1}\lambda ^{-2}\varphi \: ,\label{gen}
\end{eqnarray}

\textbf{Scalar field\begin{equation}
\varphi ''+(2+\frac{H'}{H})\varphi '+\lambda ^{-2}\varphi +\frac{1}{2}h'x+2\mu ^{2}y^{2}f_{2}\varphi =\sum \beta _{i}[1-3w_{i}-3w_{i,\rho }]\Omega _{i}\delta _{i}+\sum \Omega _{i}(1-3w_{i})\frac{\beta _{i}'}{x}\varphi \: ,\label{kgpert}\end{equation}
 }

\textbf{Metric\begin{eqnarray}
h' & = & 2\lambda ^{-2}\eta +3\sum \delta _{i}\Omega _{i}+6\varphi 'x-6\mu f_{1}y^{2}\varphi \: ,\nonumber \\
\eta ' & = & \frac{3}{2}\lambda ^{2}\sum (w_{i}+1)\Omega _{i}\theta _{i}+3\varphi x\: ,\label{eq:}\\
h'' & = & -(1+\frac{H'}{H})h'-2\left[12\varphi 'x+6\mu f_{1}y^{2}\varphi \right]-3\sum (1+3w+3w_{,\rho })\delta _{i}\Omega _{i}\: .\nonumber 
\end{eqnarray}
}

Again in view of CMB applications, it is useful to detail the adiabatic
initial conditions. The condition of zero entropy perturbations is\begin{eqnarray}
\delta S & = & \frac{\delta _{i}}{1+w_{i}}-\frac{\delta _{j}}{1+w_{j}}=0\: ,\label{eq:entropy}\\
\delta S' & = & \left(\frac{\delta _{i}}{1+w_{i}}\right)'-\left(\frac{\delta _{j}}{1+w_{j}}\right)'=0\: .\label{eq:entropy2}
\end{eqnarray}
For the scalar field\begin{equation}
\frac{\delta _{\phi }}{1+w_{\phi }}=\frac{\varphi '\phi '+\varphi a^{2}\mathcal{H}^{-2}V_{,\phi }}{\phi ^{,2}}=\frac{x\varphi '-\varphi y^{2}\mu f_{1}}{x^{2}}\: ,\label{eq:}\end{equation}
 so that applying Eqs. (\ref{eq:entropy}-\ref{eq:entropy2}) to the
scalar field and the other components we obtain the initial conditions
as \begin{eqnarray}
\varphi  & = & -\frac{x^{2}(8y^{2}\mu f_{1}\delta _{c}+2xh'+4x\delta _{c}')}{4(x^{2}+y^{2}\mu f_{1}\lambda ^{2}(6x-\beta \Omega _{c}))}\: ,\\
\varphi ' & = & -\frac{x(4x^{2}\delta _{c}-y^{2}\mu f_{1}\lambda ^{2}(-24x\delta _{c}+8\mu f_{1}y^{2}\delta _{c}+4\beta \delta _{c}\Omega _{c}+2xh'+4x\delta _{c}'))}{4(x^{2}+y^{2}\mu f_{1}\lambda ^{2}(6x-\beta \Omega _{c}))}\: .
\end{eqnarray}
In a radiation dominated era in which $\Omega _{c}\to 0$ and on super-horizon
scales ($\lambda \gg 1$), these become\begin{eqnarray}
\varphi  & = & -\frac{x(4y^{2}\mu f_{1}\delta _{c}+xh'+2x\delta _{c}')}{12y^{2}\mu f_{1}}\: ,\label{eq:}\\
\varphi ' & = & \frac{-12x\delta _{c}+4\mu f_{1}y^{2}\delta _{c}+xh'+2x\delta _{c}'}{12}\: .\label{eq:inizcond}
\end{eqnarray}
Inserting $\delta _{c}'$ from Eq. (\ref{eq:dw}) and putting initially
$\theta _{c}=0$, we can further simplify\begin{eqnarray}
\varphi  & = & -\frac{x(y^{2}\mu f_{1}+\beta x^{2})}{y^{2}\mu f_{1}}\delta _{c}\: ,\label{eq:}\\
\varphi ' & = & \frac{-3x+\mu f_{1}y^{2}}{3+\beta x}\delta _{c}\: .\label{eq:inizcond2}
\end{eqnarray}

\section{A massive dark energy field.}

Here we take a digression to consider the two effective masses of
the dark energy field, previously neglected. If $\lambda ^{2}$ is
not much larger than $\hat{m}^{2}=\hat{m}_{\phi }^{2}+\hat{m}_{\beta }^{2}$,
Eq. (\ref{phiappr-n}) in Fourier space becomes (in this section we
assume the dark energy is coupled to a single matter component, subscript
$m$, or, equivalently, that has a universal coupling to all fields)\textbf{\begin{equation}
\varphi \approx Y(k)\lambda ^{2}\beta \Omega _{m}\delta _{m},\label{phiappr-ny}\end{equation}
}where \begin{equation}
Y(k)=\frac{k^{2}}{k^{2}+a^{2}m^{2}}\: ,\label{eq:}\end{equation}
where $m=\hat{m}H.$ If we substitute in (\ref{cdm-n}) we see that
the effective potential is (neglecting the baryons)\begin{equation}
\hat{\Phi }=-\frac{3}{2}\lambda ^{2}\Omega _{m}\delta _{m}[1+\frac{4}{3}\beta ^{2}Y(k)].\label{eq:}\end{equation}
Now, let us write down the density contrast for a particle of mass
$M_{0}$ located at the origin in empty space:\begin{eqnarray}
\Omega _{m}\delta _{m} & = & \frac{\rho _{M}-\rho _{m}}{\rho _{crit}}=\frac{\kappa ^{2}M(\phi )}{3{\mathcal{H}}^{2}a}\: ,\label{eq:}\\
 &  & \nonumber 
\end{eqnarray}
where \begin{equation}
M(\phi )=M_{0}e^{-\sqrt{\frac{2\kappa ^{2}}{3}}\, \int \beta d\phi }\: ,\label{eq:}\end{equation}
where we used Eq. (\ref{eq:consmatt}). It turns out then that the
potential originated by a dark matter particle is\begin{equation}
\hat{\Phi }=-\frac{3}{2}\Omega _{m}\delta _{m}\lambda ^{2}[1+\frac{4}{3}\beta ^{2}Y(k)]=-4\pi GM(\phi )\left(\frac{1}{k^{2}}+\frac{4}{3}\beta ^{2}\frac{1}{k^{2}+a^{2}m^{2}}\right)\frac{1}{a},\label{eq:}\end{equation}
which, upon inverse Fourier transform\begin{equation}
\hat{\Phi }(r)=\frac{1}{(2\pi )^{3}}\int e^{i\mathbf{k}\cdot \mathbf{x}}\hat{\Phi }d^{3}k,\label{eq:invf}\end{equation}
becomes the Yukawa potential\begin{equation}
\hat{\Phi }(r)=-\frac{GM(\phi )}{r}\left(1+\frac{4\beta ^{2}}{3}e^{-mr}\right),\label{eq:yu1}\end{equation}
where $\mathbf{r}=a\mathbf{x}$ is the physical coordinate.

It is useful, in view of application to $N$-body simulations, to
write down explicitely the acceleration on particles. Taking Eq. (\ref{cdm-n})
and using the definition of peculiar velocity $v_{p}=adx/dt$ in terms
of the velocity used in (\ref{varia})\[
v_{p}=\mathcal{H}v\: ,\]
 we can write the acceleration equation in ordinary time $dt=ad\tau $
as \begin{equation}
\dot{v}_{p,i}=-(1-2\beta x)Hv_{p,i}-\frac{d\hat{\Phi }}{dr_{i}}.\label{finalc}\end{equation}
If we define\begin{equation}
G_{Y}(r)=G\left[1+\frac{4\beta ^{2}}{3}(1+mr)e^{-mr}\right],\label{eq:yugg}\end{equation}
we obtain the force on a dark matter particle \begin{equation}
\frac{d\hat{\Phi }}{dr_{i}}=\frac{G_{Y}(r)M(\phi )}{r^{2}}.\label{eq:forceyukawa}\end{equation}
In Eq. (\ref{finalc}) the three effects of the coupling appear clearly:
the mass $M$ depends on the time evolution of $\phi $; the gravitational
potential acquires the Yukawa correction; and the mass variation of
the test particle induces an extra friction $-2\beta x$. 

It is straightforward to generalize to the acceleration of a test
particle of type $t$ due by a distribution of several particles of
species $s$ at distances $r_{st}$ (dropping the $p$ subscript):\begin{equation}
\dot{\mathbf{v}}_{t}=-(1-2\beta _{t}x)H\mathbf{v}_{t}-\sum _{s}\frac{G_{Yst}(r_{st})M_{s}(\beta _{s},\phi )}{r_{st}^{3}}\mathbf{r}_{st},\label{finalcn}\end{equation}
where \begin{equation}
G_{Yst}(r)=G\left[1+\frac{4\beta _{s}\beta _{t}}{3}(1+mr)e^{-mr}\right].\label{eq:yug}\end{equation}
In \cite{maccio} these equations have been applied to $N$-body simulations
in the limit of constant $\beta $ and $m\to 0$.

In practice, in any dark energy model one expects the mass scale to
be much larger than the galaxy cluster scale $\approx 1$Mpc, in order
to prevent clustering, so $r\ll 1/m$ for all astrophysical scales.
However, a non-vanishing mass $m$ can have some interesting effects.

First, it is to be observed that the background dynamics depends on
the potential and its first derivative only, while the mass depends
on the second derivative. It is then possible to build viable dark
energy models that accelerate the expansion but whose mass scale $1/m$
is between a few Megaparsecs and $H_{0}^{-1}=3000$ Mpc/$h$; in this
case the perturbations of the scalar field would be directly observable
through, e.g., weak lensing. 

Second, while the mass $m_{\phi }$ depends exclusively on the potential
$V(\phi )$, the coupling mass\begin{equation}
m_{\beta }^{2}\equiv x^{-1}\Omega _{m}\beta '_{m}H^{2}=\sqrt{\frac{6}{\kappa ^{2}}}\Omega _{m}\beta _{,\phi }H^{2}\: ,\label{eq:couplingmass}\end{equation}
 depends on the matter content $\Omega _{m}$ . In a inhomogeneous
background, we can expect that $m_{\beta }^{2}$ will be proportional
to $\rho _{m}$. This raises an interesting question, recently posed
in the {}``chameleon'' model of Ref. \cite{khoury} and also in
the context of $\alpha $-varying models \cite{mb}: is it possible
to have a scalar field with a large mass near a massive body like
the Earth and a very low one in space ? This would open the possibility
that scalar gravity escapes detection on Earth laboratories, where
the Yukawa term would be exponentially suppressed if $1/m$ is on
the submillimetric scale, even if $\beta $ were of order unity (here
we neglect any possible upper bound from cosmology). In \cite{khoury}
it has been hypothesized that a model with a constant $\beta $ do
in fact contain a density-dependent mass but our calculations show
clearly that $\beta '\not =0$ is a necessary condition. This possibility
will be discussed in another paper.

\section{Second order equations}

Here we extend the previous calculations to second order in the Newtonian
regime and in the non-relativistic limit. This will allow to evaluate
higher order moments of the gravitational clustering. Higher order
perturbation equations in a varying dark matter mass scenario and
in scalar-tensor theories which may be reduced to particular cases
of the present model have been studied in \cite{nonlin}. 

Here for simplification we consider a single matter fluid with $w=0$.
In the non-relativistic limit the quadrivelocity remains first order\begin{equation}
u^{\alpha }=\frac{dx^{\alpha }}{ds}\approx \frac{dx^{\alpha }}{\sqrt{g_{00}}d\tau }.\label{eq:}\end{equation}
 The general conservation equations at second order are then ($v^{2}=v_{i}v^{i}$)
\textbf{\begin{eqnarray}
\delta '+\nabla _{i}(1+\delta )v_{i} & = & 3(1+\delta )\Phi '-2\beta \varphi '(1+\delta )+6\Phi \Phi '-v_{i}\nabla _{i}(\Psi -3\Phi )-H^{2}v^{2}(1-2\beta x)\nonumber \\
-2\beta '\varphi (1+\delta +\frac{\varphi '}{x})-\frac{\varphi ^{2}}{x}(\beta ''-\beta '\frac{\phi ''}{\phi '})\: , &  & \nonumber \\
(1+\delta -2\Psi -2\Phi )\left[v_{i}'+\left(1+\frac{\mathcal{H}'}{\mathcal{H}}-2\beta x\right)v_{i}\right] & = & -\frac{1+\delta }{\mathcal{H}^{2}}\nabla _{i}(\Phi -2\beta \varphi )+2\Psi \frac{\nabla _{i}\Psi }{\mathcal{H}^{2}}-v_{j}\nabla _{j}v_{i}+v_{i}(2\Phi '+\Psi ')\label{eq:gensec}\\
+2\beta '\frac{\varphi }{x\mathcal{H}^{2}}\nabla _{i}\varphi \: , &  & \nonumber 
\end{eqnarray}
}where \textbf{}the \textbf{}second line of each equation contains
the terms from the variation of $\beta (\phi )$. The scalar field
equation is\begin{eqnarray}
[\varphi ''+\big (2+\frac{\mathcal{H}'}{\mathcal{H}}\big )\varphi '-2y^{2}\mu f_{1}\Psi ](1-2\Psi ) & + & \nonumber \\
2y^{2}\mu ^{2}\varphi (f_{2}+\mu f_{3}\varphi )+\varphi '(3\Phi '+\Psi ')-\Phi 'x(3+6\Phi -6\Psi )-\Psi 'x(1-4\Psi ) & - & \nonumber \\
(1+2\Phi )\frac{\nabla ^{2}}{\mathcal{H}^{2}}\varphi +\frac{1}{\mathcal{H}^{2}}(\nabla _{i}\varphi )\nabla _{i}(\Phi -\Psi ) & = & \label{eq:pert5}\\
\beta \Omega _{m}[2\Psi (1-2\Psi )+\delta -\mathcal{H}^{2}v^{2}] & + & \nonumber \\
\frac{\varphi }{x}\Omega _{m}\beta '(1+\delta _{c})+\frac{\varphi ^{2}}{2x^{2}}\Omega _{m}(\beta ''-\beta '\frac{\phi ''}{\phi '}) &  & \nonumber 
\end{eqnarray}
where $f_{3}$ is defined in Eq. (\ref{eq:fn}).

Let us now derive the Newtonian limit. We can use the metric equations
at first order, in particular the relation $\Phi =\Psi $ . As before,
$\Phi $ and $\varphi $ are of order $\lambda ^{2}$ (with respect
to $\delta $) and we neglect the time derivatives of $\varphi $.
We obtain from (\ref{eq:gensec})\begin{eqnarray}
\delta '_{c}+\nabla _{i}(1+\delta )v_{i} & = & 2v_{i}\nabla _{i}\Phi -H^{2}v^{2}(1-2\beta x)\: ,\nonumber \\
v_{i}'+\left(1+\frac{\mathcal{H}'}{\mathcal{H}}-2\beta x\right)v_{i} & = & -\frac{1}{\mathcal{H}^{2}}\nabla _{i}(\Phi -2\beta \varphi )-\frac{1}{1+\delta }v_{j}\nabla _{j}v_{i}\: .\label{eq:pert6}
\end{eqnarray}
Furthermore, in the first equation we can neglect the term $H^{2}v^{2}(1-2\beta x)$
because of the non-relativistic approximation and the term $2v_{i}\nabla _{i}\Phi $
because both $v_{i}$ and $\nabla _{i}\Phi $ are of order $\lambda $
at first order. Finally, in the second equation we can approximate
$1+\delta \approx 1$ at the denominator in the last term, since it
gives a third order correction. Therefore we are left with \begin{eqnarray}
\delta '_{c}+\nabla _{i}(1+\delta )v_{i} & = & 0\: ,\label{eq:}\\
v_{i}'+\left(1+\frac{\mathcal{H}'}{\mathcal{H}}-2\beta x\right)v_{i} & = & -\frac{1}{\mathcal{H}^{2}}\nabla _{i}(\Phi -2\beta \varphi )-v_{j}\nabla _{j}v_{i}\: .\label{eq:pert8}
\end{eqnarray}
 Applying the same approximations, for the scalar field we obtain\begin{eqnarray}
\varphi ''+\big (2+\frac{\mathcal{H}'}{\mathcal{H}}\big )\varphi '-2y^{2}\mu f_{1}\Phi -2\beta \Omega _{m}\Phi  &  & \nonumber \\
-\frac{\nabla ^{2}}{\mathcal{H}^{2}}\varphi -4\Phi 'x+2y^{2}\mu ^{2}f_{2}\varphi +2y^{2}\mu ^{3}f_{3}\varphi ^{2} & = & \beta \Omega _{m}\delta \: ,\label{pert7}
\end{eqnarray}
which neglecting the time derivatives of $\varphi $ reduces to the
non-linear Klein-Gordon equation\begin{equation}
\nabla ^{2}\varphi -m^{2}\varphi -\sigma \varphi ^{2}=-\beta \Omega _{m}\delta \mathcal{H}^{2}\: ,\label{eq:kg3}\end{equation}
where we defined the non-linear correction\begin{equation}
\sigma =2y^{2}\mu ^{3}f_{3}H^{2}\: .\label{eq:m3}\end{equation}
If $m$ and $\sigma $ are negligible with respect to the lenght scale
$\lambda $ then we see that the non-linear conservation equations
in coupled dark energy coincide with the usual non-linear Newtonian
perturbation equations with an effective potential $\hat{\Phi }$
and a correction in the Euler equation due to the time variation of
the dark matter mass :\begin{eqnarray}
\delta '+\nabla _{i}(1+\delta )v_{i} & = & 0\: ,\label{eq:98}\\
v_{i}'+(1+\frac{\mathcal{H}'}{\mathcal{H}}-2\beta x)v_{i}+v_{j}\nabla _{j}v_{i} & = & -\frac{1}{\mathcal{H}^{2}}\nabla _{i}\hat{\Phi }\: ,\label{eq:99}\\
\nabla ^{2}\hat{\Phi } & = & 4\pi G_{mm}\rho _{m}\delta \: .\label{eq:iniz-comp}
\end{eqnarray}
If $\sigma \varphi ^{2}$ is negligible but $m\varphi $ is not then
one should use the Yukawa correction to $G_{mm}$ (in this case one
should also assume that the coefficients of the terms in $\Phi ,\Phi '$
in (\ref{eq:kg3}) are smaller than $\hat{m}^{2}$). As it has been
shown, Eqs. (\ref{eq:98}-\ref{eq:iniz-comp}) are valid also when
$\beta $ is a function of $\phi $.

\section{Conclusions}

This paper is meant to set up the formalism for future work on non-linear
properties of dark energy, with an emphasis on its coupling to matter.
Along this survey we found several results that we summarize here.

\emph{a}) We derived the general equations for the linear pertubation
growth for a general dark energy potential and a general species-dependent
interaction with matter. This generalize previous work.

\emph{b}) We derived an analytical relation between the bias induced
by a species-dependent coupling and the growth exponent of the linear
perturbations, as well as their values in term of the fundamental
parameters in the case of exponential potential.

\emph{c}) We discussed the Yukawa correction to the gravitational
potential due to dark energy interaction and we found that a density-dependent
effective dark energy mass arises only if the coupling is non-constant.
The consequence of this effect on equivalence principle experiments
will be discussed in another paper.

\emph{d}) We derived the second-order perturbation equations in coupled
dark energy and their Newtonian limit. We showed that the coupling
introduces three corrections to the standard Newtonian fluid equations,
one proportional to the velocity and the others which can be absorbed
in the gravitational potential. These equations wil be used in a subsequent
paper to derive the large scale skewness of coupled dark energy.

\begin{acknowledgments}
I acknowledge useful discussions with S. Bonometto, J. Khoury, A.
Macciò, R. Mainini, D. Mota, F. Perrotta, C. Quercellini, N. Sakai,
and D. Tocchini-Valentini.
\end{acknowledgments}

\end{document}